\documentclass[twocolumn,showpacs,preprintnumbers,amsmath,amssymb,superscriptaddress]{revtex4-1}
\usepackage{graphicx}
\usepackage{dcolumn}
\usepackage{bm}
\usepackage{CJK, CJKnumb}
\usepackage{color}
\begin{document}
\begin{CJK*}{GBK}{song}

\title{Cone-dependent retro and specular Andreev reflections in AA-stacked bilayer graphene}
\author{Wei-Tao Lu} \email{physlu@163.com}
\affiliation{School of Sciences, Nantong University, Nantong 226019, China}
\author{Qing-Feng Sun} \email{sunqf@pku.edu.cn}
\affiliation{International Center for Quantum Materials, School of Physics, Peking University, Beijing 100871, China}
\affiliation{Hefei National Laboratory, Hefei 230088, China}

\begin{abstract}
We theoretically study the Andreev reflection (AR) in AA-stacked bilayer graphene/superconductor (AABG/SC) junction.
AABG has a linear gapless energy band with two shifted Dirac cones and the electronic states are described by the cone indices.
The results indicate that the property of AR strongly depends on the cone degree of freedom.
In the absence of the inter-layer potential difference, only intra-cone AR and normal reflection (NR) could occur, and the inter-cone process is forbidden.
By adjusting the potential, the intra-cone AR can be specular AR (SAR) in one cone and it is retro-AR (RAR) in the other cone.
The existence of the inter-layer potential difference would lead to the inter-cone scattering.
As a result, double ARs and double NRs can take place between the two cones.
The inter-cone SAR could happen in a broad potential region.
Furthermore, the inter-cone retro-NR (RNR) could happen as well.
The switch between SAR and RAR, and the switch between specular NR (SNR) and RNR can be achieved by regulating the potential.
Therefore, different cone carriers can be separated spatially based on the RAR and SAR.
The cone-dependent Andreev conductance may be separately measured near the critical values where RAR crosses over to SAR. 
\end{abstract}
\maketitle

\section{Introduction}

Andreev reflection (AR) is an important phenomenon of quantum transport
which occurs at the interface between the metal and superconductor (SC) \cite{Andreev}.
An incident electron from the metal is reflected as a hole at
the metal/SC interface and a Cooper pair is formed in the SC.
In conventional metal, the hole is reflected back along the path
of incident electron which is called retro-AR (RAR).
However, the hole can be expected to be reflected specularly
at the graphene/SC interface due to the interband Andreev processes,
which is named specular AR (SAR) \cite{Beenakker, Beenakker2}.
Recently, the SAR process has been discovered and studied in graphene \cite{Beenakker2, Greenbaum, ZhangQ, ChengSG, addref1, Komatsu, Efetov, Sahu, addref2}, silicene \cite{LiH, LuWT}, semimetals \cite{HouZ, Azizi, ChengQ, ChengQ2}, topological insulators \cite{Majidi}, and two-dimensional electron gas \cite{LvB}.
Because of the quantum interference of the reflected holes from two superconductor terminals, the RAR and SAR processes could be selected by tuning the phase difference \cite{ChengSG}.
Experimentally, the observation of transitions between RAR to SAR has been reported at the van der Waals interface of graphene and NbSe$_2$ superconductor \cite{Efetov, Sahu}.

Bilayer graphene (BG) exhibits additional properties that make it distinct from monolayer graphene, such as trigonal warping \cite{Cserti} and unusual quantum Hall effect \cite{McCann}.
BG exists in three configurations: AA stacking, AB stacking, and twisted bilayer \cite{Rozhkov}.
For the usual AB-stacked BG (ABBG), the A sublattice of the top layer
is stacked directly above the B sublattice of the bottom layer.
ABBG can open a gap between the conduction and valence bands
by adjusting the gate voltages between the two layers \cite{Ohta, Castro, addref3, LuWT2}.
ABBG is also a candidate to observe the crossover from RAR to SAR
since the Fermi energy broadening near the Dirac point is weaker
compared to monolayer graphene \cite{Efetov2, Soori}.
Some interesting features of AR in ABBG/SC hybrid system have been theoretically
investigated by several groups \cite{Ludwig, Schroer, Efetov2, Soori, QiF, WuX}.
The differential conductance across the ABBG/SC junction suggests a characteristic signature of the crossover from intraband RAR to interband SAR that manifests itself in a strongly suppressed interfacial conductance \cite{Efetov2}.
However, the signature of SAR is not very pronounced in experiment \cite{Efetov}.
It is found that the SAR process can be enhanced in the presence of a Zeeman field due to the separation of Dirac points for up spin and down spin \cite{Soori}.
Furthermore, the valley-entangled Cooper pair splitter could be realized and controlled based on the nonlocal AR in the ABBG/SC/ABBG junction \cite{Schroer, QiF, WuX}.

For the AA-stacked bilayer graphene (AABG), each carbon atom of the top layer is stacked directly above the corresponding atom of the bottom layer \cite{Rozhkov}.
The stability and electronic structure of AABG have been theoretically predicted \cite{Andres}, and stable samples of AABG have been realized in experiment \cite{LeeJK, LiuZ}.
Contrary to ABBG, AABG possesses a linear gapless energy spectrum with two shifted Dirac cones in the low energy regime.
Recently, the electronic properties of AABG have attracted considerable attention \cite{HsuYF, Lofwander, Dyrdal, HoYH, WangD, LinCY, Qasem, Rakhmanov, Akzyanov, Apinyan, Gonzalez, Tabert, WangD2, Sanderson, Abdullah, Abdullah2, ChenRB, ChiuCW, LiY}, including quantum Hall effect \cite{HsuYF, Lofwander, Dyrdal}, Landau levels \cite{HoYH, WangD, LinCY}, antiferromagnetic states \cite{Rakhmanov, Akzyanov, Apinyan}, and electronic transport \cite{Gonzalez, Tabert, WangD2, Sanderson, Abdullah, Abdullah2}.
Since the energy bands of the AABG are different from ABBG, many interesting characteristics of quantum Hall effect in AABG are predicted, which are not seen in ABBG \cite{HsuYF}.
The magneto-optical absorption spectra of AABG exhibit two kinds of absorption peaks resulting from two groups of Landau levels, different from the spectra of ABBG \cite{HoYH}.
Discussion on the Landauer conductance for both AABG and ABBG demonstrates
that the conductance is very sensitive to the geometry of the system,
which could be used as an electromechanical switch \cite{Gonzalez}.
The electrons in AABG are not only chiral but also are described by a cone degree of freedom.
The unique cone transport, together with the negative refraction,
suggests the possibility of cone-tronic devices based on AABG \cite{Sanderson}.
The van der Waals domain wall between monolayer graphene and AABG which is described by a local variation of interlayer coupling,
can be used to generate two distinct types of collimated electron beams that correspond to the lower and upper cones in AABG \cite{Abdullah2}.
Nonetheless, the AR phenomenon in the AABG still need to be explored,
which may provide specific signatures for the AABG.

In this work, we study the property of AR in the AABG/SC junction
in the framework of the Bogoliubov-de Gennes equation.
The cone-dependent RAR and SAR are found in AABG,
which are not observed in ABBG and monolayer graphene.
In the presence of the inter-layer potential difference, both the intra-cone and inter-cone scatterings can take place,
giving rise to the double ARs and double NRs.
The inter-cone AR (or NR) may be SAR and RAR (or SNR and RNR)
depending on the value of the potential.
The Andreev conductances for the upper cone and lower cone exhibit different features, and they can be separately measured near the critical values for SAR and RAR.
The property of AR and NR scatterings may be understood by the ray equations
and the orthogonality of wave functions.

The rest of this paper is organized as follows. The Hamiltonian of the AABG/SC junction and the reflection processes at the two cones are given and studied in Sec. II. The results on cone-polarized AR and double ARs are discussed in Sec. III. A brief summary is presented in Sec. IV.

\section{Theoretical Formulation}
\subsection{Theoretical model and reflection probabilities}

The AABG consists of two graphene layers and each layer consists of two triangular sublattices, $A$ and $B$. Considering the nearest-neighbor hopping, the system can be described by the tight-binding Hamiltonian \cite{Rakhmanov},
\begin{align}
\mathcal{H} = &-t \sum_{\langle i j\rangle \ell,\sigma} a_{i \ell \sigma}^\dag a_{j \ell \sigma} \nonumber \\
&- \gamma \sum_{i\in A ,\sigma}  a_{i1\sigma}^\dag a_{i2\sigma}
- \gamma  \sum_{j\in B ,\sigma} a_{j1\sigma}^\dag a_{j2\sigma} +H.c.,
\end{align}
where $a_{i \ell \sigma}^\dag$ ($a_{i \ell \sigma}$) is the creation (annihilation) operator with spin $\sigma$ at the sublattice $i \in A, B$ in the layer $\ell=1,2$.
The first term is the in-plane nearest-neighbor hopping with amplitude $t\approx 2.8 eV$.
The second and third terms describe the inter-layer nearest-neighbor hopping with amplitude $\gamma \approx 200 meV$ \cite{Tabert}.
The in-plane and inter-layer next-nearest neighbor hoppings are very weak compared with $t$ and $\gamma$, which are neglected in the following calculation.
In the basis $\psi=(\psi_{A1},\psi_{B1},\psi_{A2},\psi_{B2})^T$, the tight-binding Hamiltonian in the $k$ space can be written in the following matrix representation:
\begin{align}
H_{AABG}^\eta = \left(\begin{array}{cc} H_\eta  +U \tau_0 + \delta \tau_0  &  \gamma \tau_0  \\  \gamma \tau_0  &  H_\eta + U\tau_0 - \delta\tau_0  \end{array}\right),
\end{align}
with $H_\eta=\hbar v_F (k_x \tau_x + \eta k_y \tau_y)$.
$\tau=(\tau_x,\tau_y)$ are the Pauli matrices in the sublattice $A$ and $B$ spaces
and $\tau_0$ is unit matrix.
The valley index $\eta=\pm 1$ corresponds to the $K$ and $K'$ valleys.
The inter-layer potential difference is $\delta=(U_1-U_2)/2$ and the potential is $U=(U_1+U_2)/2$ with $U_1$ and $U_2$ being the electrostatic potential at the two layers, which can be induced by the top and back gates on the sample.
$\delta$ could open a band gap in ABBG due to the asymmetric inter-layer coupling \cite{Ohta, Castro}. However, $\delta$ cannot open a gap in AABG since the coupling is symmetric, but it could induce the inter-cone scattering, as discussed in Sec. III (B).

For the proposed AABG/SC junction in the $(x,y)$ plane, the SC electrode covers the region $x>0$ and the normal AABG electrode covers the region $x<0$.
The inter-layer potential difference $\delta$ and the potential $U$ are only applied in the AABG region. 
Note that here the SC refers to superconducting AABG, which can be induced via the proximity effect by depositing SC on AABG \cite{FMiao, Heersche, Choi}. 
The electron and hole excitations can be described by the Bogoliubov-de Gennes (BdG) Hamiltonian
\begin{align}
H_{BdG} = \left(\begin{array}{cc}  H_{AABG}^\eta  &  \Delta \textbf{I}_{4\times4}  \\  \Delta \textbf{I}_{4\times4}  &  - \mathcal{T}  H_{AABG}^\eta  \mathcal{T}^{-1}  \end{array}\right),
\end{align}
where $\textbf{I}_{4\times4}$ is the $4\times4$ unit matrix and $\Delta$ is the superconducting gap coupling the time-reversed electron and hole states.
$\mathcal{T}= \begin{pmatrix} \tau_z  &  0  \\  0  &  \tau_z  \end{pmatrix} \mathcal{C}$
with the complex conjugation $\mathcal{C}$ is the time-reversal operator, which satisfies $\mathcal{T}^{-1}=\mathcal{T}$ and $\mathcal{T} H_{AABG}^{\eta} \mathcal{T}^{-1} = H_{AABG}^{-\eta}$.
The Hamiltonian $H_{BdG}$ is time-reversal invariant and the two valleys are degenerate. Thus, we only discuss the AR process of one valley in the following.

The dispersion relation for electron and hole in the AABG region can be written as
\begin{eqnarray}
\varepsilon_{e} &= &\pm \hbar v_F \sqrt{k_x^2+k_y^2} + \alpha \sqrt{\gamma^2+\delta^2} + U, \label{Ee}\\
\varepsilon_{h} &= &\pm \hbar v_F \sqrt{k_x^2+k_y^2} - \alpha \sqrt{\gamma^2+\delta^2} - U, \label{Eh}
\end{eqnarray}
and $\alpha=\pm$ for the upper and lower cones. In the SC region, the eigenvalue for Hamiltonian $H_{BdG}$ has the form
\begin{align}
\varepsilon_S = \pm \sqrt{(\hbar v_F \sqrt{k_x^2+k_y^2} \pm \gamma)^2 + \Delta^2}.
\end{align}
Fig. 1(a) shows the band structure in the AABG region calculated by
Eqs. (\ref{Ee}) and (\ref{Eh}). The solid and dashed curves are
energy bands for electron and hole, respectively.
We can find that the bands of pristine AABG are two copies of the band of monolayer graphene, one shifted by $-\gamma$ and the other by $+\gamma$.
Thus, the linear band consists of two Dirac cones, that is the lower and upper cones [see black and red cones in Fig. 1(a)].

The eigenvectors of the junction are given by solving the BdG Hamiltonian (see Appendix for detail).
For an electron injecting from the upper cone, the wave functions in the
AABG ($x<0$) and SC ($x>0$) regions of the AABG/SC junction can be written as:
\begin{align}
& \psi_{AABG} (x) = \psi_{e1}^{+} + r^+_{e1}\psi_{e1}^{-} + r^+_{e2}\psi_{e2}^{-} + r^+_{h1}\psi_{h1}^{-} + r^+_{h2}\psi_{h2}^{-},    \\
& \psi_{SC} (x) = t^+_{S1}\psi_{S1}^{+} + t^+_{S2}\psi_{S2}^{+} + t^+_{S3}\psi_{S3}^{+} + t^+_{S4}\psi_{S4}^{+}.
\end{align}
For the eigenvectors $\psi_{e(h)1}^{\pm}$ and $\psi_{e(h)2}^{\pm}$, the indexes $1$ and $2$ refer to the upper and lower cones, $e$ and $h$ refer to electron and hole, ${\pm}$ represent the right-going and left-going propagating waves, respectively.
$\psi_{S1 \texttt{-} S4}$ are the eigenvectors in the SC region (see Appendix for detail).
$r^+_{e1(e2)}$, $r^+_{h1(h2)}$, and $t^+_{S1\texttt{-}S4}$ are
the NR coefficient, AR coefficient, and tunneling coefficient, respectively.
Based on the continuity of wave function and the conservation of current density at the interface with $x=0$, the reflection coefficients can be derived, and then the reflection probabilities can be obtained as
\begin{align}
& R^{++}_{N} = |r^+_{e1}|^2,   \\
& R^{+-}_{N} =  \frac{E-U-\Gamma}{E-U+\Gamma} Re \left(\frac{k_{e2x}}{k_{e1x}}\right)   |r^+_{e2}|^2,    \\
& R^{++}_{A} =  \frac{E-U-\Gamma}{E+U+\Gamma} Re \left(\frac{k_{h1x}}{k_{e1x}}\right)   |r^+_{h1}|^2,    \\
& R^{+-}_{A} =  \frac{E-U-\Gamma}{E+U-\Gamma} Re \left(\frac{k_{h2x}}{k_{e1x}}\right)   |r^+_{h2}|^2,
\end{align}
where $\Gamma=\sqrt{\gamma^2+\delta^2}$ and $E$ is the energy of the incident electron.
$R^{++}_{N}$ and $R^{++}_{A}$ are the intra-cone NR and AR for the upper cone. $R^{+-}_{N}$ and $R^{+-}_{A}$ are the inter-cone NR and AR for electron from the upper cone ($+$) to the lower cone ($-$).
$k_{e1(2)x}$ and $k_{h1(2)x}$ are $x$ components of momentums $k_{e1(2)}$ and $k_{h1(2)}$, respectively.
In the same way, when electron injects from the lower cone, the reflection probabilities can be derived as
\begin{align}
& R^{-+}_{N} =  \frac{E-U+\Gamma}{E-U-\Gamma} Re \left(\frac{k_{e1x}}{k_{e2x}}\right)   |r^-_{e1}|^2,   \\
& R^{--}_{N} =  |r^-_{e2}|^2,    \\
& R^{-+}_{A} =  \frac{E-U+\Gamma}{E+U+\Gamma} Re \left(\frac{k_{h1x}}{k_{e2x}}\right)    |r^-_{h1}|^2,    \\
& R^{--}_{A} =  \frac{E-U+\Gamma}{E+U-\Gamma} Re \left(\frac{k_{h2x}}{k_{e2x}}\right)  |r^-_{h2}|^2.
\end{align}
Here, $R^{--}_{N}$ and $R^{--}_{A}$ are the intra-cone NR and AR for the lower cone. $R^{-+}_{N}$ and $R^{-+}_{A}$ are the inter-cone NR and AR for electron from the lower cone ($-$) to the upper cone ($+$).
The defined reflection probabilities satisfy the conservation conditions,
\begin{align}
R^{\pm}_{total}=R^{\pm +}_{N} + R^{\pm -}_{N} + R^{\pm +}_{A} + R^{\pm -}_{A} =  1,
\end{align}
in the subgap with the energy $|E|<\Delta$.

The Andreev conductance for the electron from the upper and lower cones at zero temperature can be evaluated by the Blonder-Tinkham-Klapwijk formula \cite{Blonder}:
\begin{align}
G^{\pm}_{AR}= G_0 \int_0^{\pi/2} (1+R^{\pm+}_{A}+R^{\pm-}_{A}-R^{\pm+}_{N}-R^{\pm-}_{N}) \cos \theta d \theta,
\end{align}
where $\theta$ is the incident angle of electron with respect to the $x$ direction, $G_0=e^2 N^{\pm}_0(E)/h$ characterizes the ballistic conductance of the AABG/SC junction, $N^{\pm}_0(E)=W |k_{e1(2)}| / \pi$ denotes the number of transverse modes, $W$ labels the width of the junction, and the incident
energy $E=eV_b$ with the voltage $V_b$ on the AABG/SC junction.

\subsection{Reflection processes at the two cones}

\begin{figure}
\includegraphics[width=8.0cm,height=7.0cm]{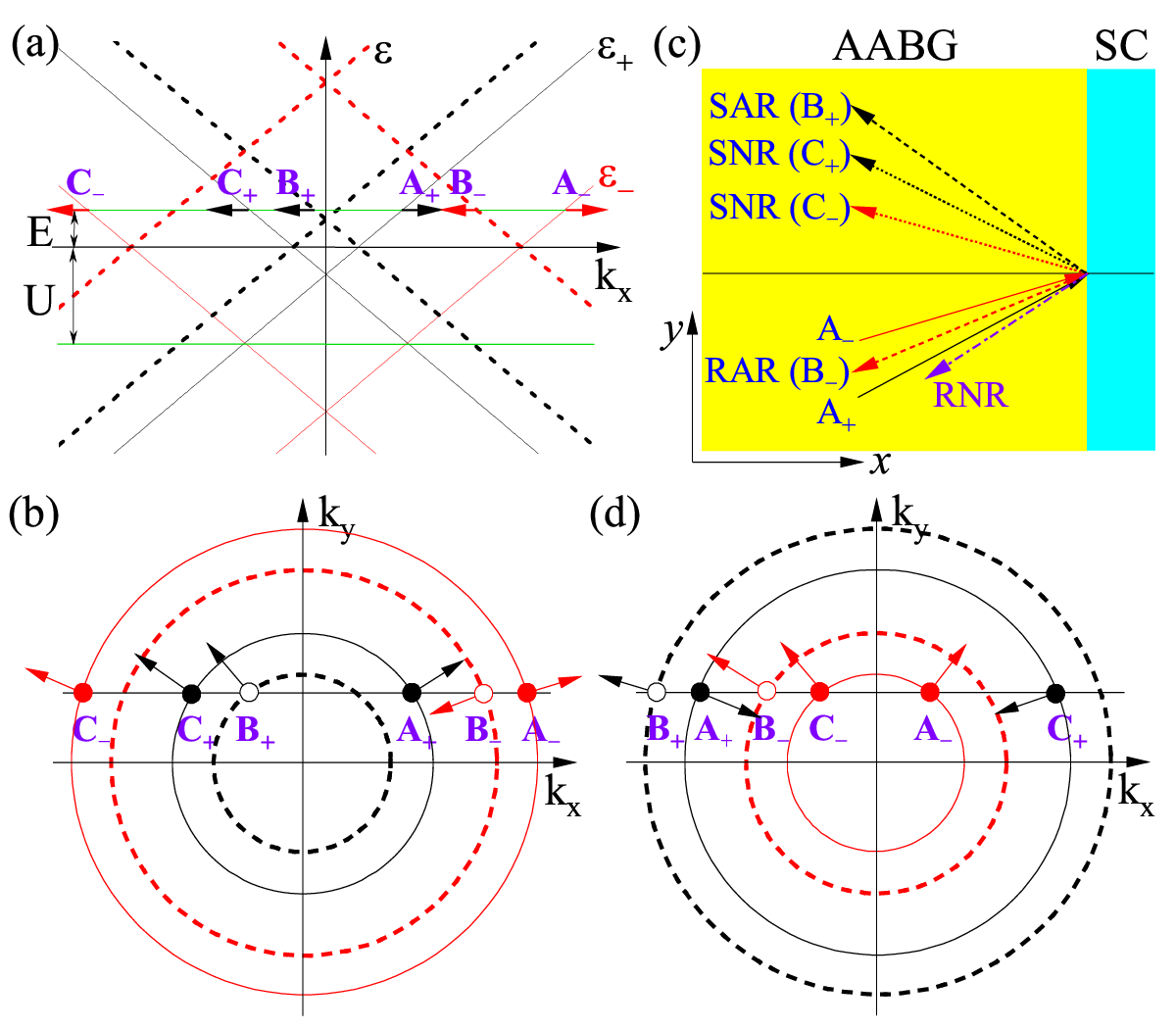}
\caption{ (a) Band structures in the AABG region. An incident electron from the $A_\pm$ points can be Andreev reflected as a hole from $B_\pm$ points and normally reflected as an electron from $C_\pm$ points.
The solid and dashed curves are the energy bands for electron and hole.
The black and red curves are the energy bands at upper and lower cones, respectively.
(b) and (c) The projection of the Andreev scattering and normal scattering processes (b) on the $k_x-k_y$ plane and (c) on the $x-y$ real space.
The black and red arrows denote the group velocities at the upper and lower cones. 
The arrows with filled circles and empty circles are for the electrons and holes, respectively.
The potential satisfies $U_{c1}<U<U_{c2}$ in (a-c).
(d) The Andreev scattering and normal scattering processes on the $k_x-k_y$ plane when $U_{c3}<U<U_{c4}$. }
\end{figure}

Before giving the numerical results, in this subsection, we discuss the property of AR and NR in the scattering process by analyzing the eigenvectors and the ray equations.
For a given incident energy $E$,
there are two incident electronic states corresponding to the wave vectors $A_+$ and $A_-$ and four reflection states corresponding to the wave vectors $B_+$, $B_-$, $C_+$, and $C_-$, as shown Fig. 1(a).
The two ARs correspond to the wave vectors $B_+$ and $B_-$. The two NRs correspond to the wave vectors $C_+$ and $C_-$.
$A_+$, $B_+$, and $C_+$ are the wave vectors at upper cone while $A_-$, $B_-$, and $C_-$ are the wave vectors at lower cone.
Below we discuss the effect of the inter-layer potential difference $\delta$ and the potential $U$ on the AR and NR processes.

When $\delta=0$, regardless of the values of $U$, the eigenvectors in the AABG region satisfy the relations
\begin{align}
\langle \varphi^{\pm}_{ei}| \varphi^{\pm}_{ej} \rangle = 0,  \hspace{5mm}  \langle \varphi^{\pm}_{ei}| \varphi^{\pm}_{hj} \rangle = 0,
\end{align}
with $i,j=1,2$ and $i \neq j$. That is, the eigenvectors with different cone indices are orthometric.
The orthogonality of the wave function suggests that both AR and NR
between the cones are forbidden. The AR and NR are only allowed in the intra-cone.

When $\delta \neq 0$ and $U=0$, the eigenvectors satisfy
\begin{align}
\langle \varphi^{\pm}_{ei}| \varphi^{\pm}_{ej} \rangle \neq 0,  \hspace{5mm}  \langle \varphi^{\pm}_{ei}| \varphi^{\pm}_{hj} \rangle = 0,
\end{align}
suggesting that NR between the cones is allowed but the AR between the cones is still forbidden.
Consequently, not only the intra-cone NR but also the inter-cone NR would occur, and it is referred as double NRs.

When $\delta \neq 0$ and $U \neq 0$, the eigenvectors satisfy
\begin{align}
\langle \varphi^{\pm}_{ei}| \varphi^{\pm}_{ej} \rangle \neq 0,  \hspace{5mm}  \langle \varphi^{\pm}_{ei}| \varphi^{\pm}_{hj} \rangle \neq 0,
\end{align}
suggesting that both NR and AR between the cones are allowed.
The intra-cone AR and inter-cone AR can occur.
This means that both double NRs and double ARs could occur.
Therefore, the occurrence of AR and NR strongly depends on the inter-layer potential difference $\delta$ and the potential $U$.

For the AR process, both SAR and RAR processes could take place in the AABG/SC junction.
Next, we analyze the SAR and RAR by the ray equations \cite{Landau}, which can be controlled by $\delta$ and $U$.
Based on the dispersion relations in Eqs. (\ref{Ee}) and (\ref{Eh}),
there are four critical values for the potential in the AR process:
\begin{align}
& U_{c1(c2)}=-\Gamma-(+)E,  \\
& U_{c3(c4)}=\Gamma-(+)E.
\end{align}
When $U_{c1}<U<U_{c2}$, as shown in Fig. 1(a), the wave vector $\textbf{k}$ and the group velocity $\textbf{v}$ for the incident electron at the $A_+$ and $A_-$ points are positive, and they have the relation:
\begin{align}
\textbf{k}_{(A_+)} \cdot \textbf{v}_{(A_+)} > 0,  \hspace{5mm}  \textbf{k}_{(A_-)} \cdot \textbf{v}_{(A_-)} > 0.
\end{align}
However, for the Andreev reflected holes at the $B_+$ and $B_-$ points, $\textbf{k}$ and $\textbf{v}$ satisfy
\begin{align}
\textbf{k}_{(B_+)} \cdot \textbf{v}_{(B_+)} > 0,  \hspace{5mm}  \textbf{k}_{(B_-)} \cdot \textbf{v}_{(B_-)} < 0.
\end{align}
For the normal reflected electrons at $C_+$ and $C_-$ points, $\textbf{k}$ and $\textbf{v}$ satisfy
\begin{align}
\textbf{k}_{(C_+)} \cdot \textbf{v}_{(C_+)} > 0,  \hspace{5mm}  \textbf{k}_{(C_-)} \cdot \textbf{v}_{(C_-)} > 0.
\end{align}
Comparing the ray equations (24) and (26), one may find that $\textbf{k}_{(C_\pm)} \cdot \textbf{v}_{(C_\pm)}$ of the reflected electron has the same sign as that of the incident electron, suggesting that the reflection of electron at both $C_+$ and $C_-$ points is specular reflection.
Nevertheless, the ray equations (24) and (25) indicate that $\textbf{k}_{(B_+)} \cdot \textbf{v}_{(B_+)}$ has the same sign as $\textbf{k}_{(A_+)} \cdot \textbf{v}_{(A_+)}$ and $\textbf{k}_{(A_-)} \cdot \textbf{v}_{(A_-)}$
and so the reflection of hole at $B_+$ point is SAR regardless of the inter-cone
and intra-cone ARs.
While $\textbf{k}_{(B_-)} \cdot \textbf{v}_{(B_-)}$ has the opposite sign as $\textbf{k}_{(A_+)} \cdot \textbf{v}_{(A_+)}$ and
$\textbf{k}_{(A_-)} \cdot \textbf{v}_{(A_-)}$,
so the reflection at $B_-$ point is RAR [see Figs. 1(b) and 1(c)].
On the other hand, when $U_{c3}<U<U_{c4}$ in Fig. 1(d),
the intra-cone AR in the upper cone and
the inter-cone AR from the upper cone to the lower cone are RAR,
but the intra-cone AR in the lower cone
and the inter-cone AR from the lower cone to the upper cone are SAR.
Furthermore, it is possible that the SAR and RAR processes could happen at the same time, as discussed in the following.

\section{Results and discussions}
In this section we discuss the property of AR process in the AABG/SC junction and its dependance on cone degree of freedom, including the cone-polarized AR at $\delta=0$ in subsection III (A) and the double ARs at $\delta\neq0$ in subsection III (B).
The value of superconducting gap is set as $\Delta=0.01\gamma$ and its value does not affect the conclusion. $\Delta$ is the unit of $E$, $\delta$, and $U$.

\subsection{Cone-polarized Andreev reflection}

First, we discuss the AR process in the absence of $\delta$ and the results are given in Figs. 2-4.
Equation (19) indicates that when $\delta=0$ the AR and NR for the inter-cone are forbidden and only the intra-cone processes are permitted.
For the electron from upper cone, $R^{++}_{N}+R^{++}_{A}=1$
and $R^{+-}_{N}=R^{+-}_{A}=0$.
For the electron from lower cone, $R^{--}_{N}+R^{--}_{A}=1$
and $R^{-+}_{N}=R^{-+}_{A}=0$.

\begin{figure}
\includegraphics[width=8.0cm,height=5.0cm]{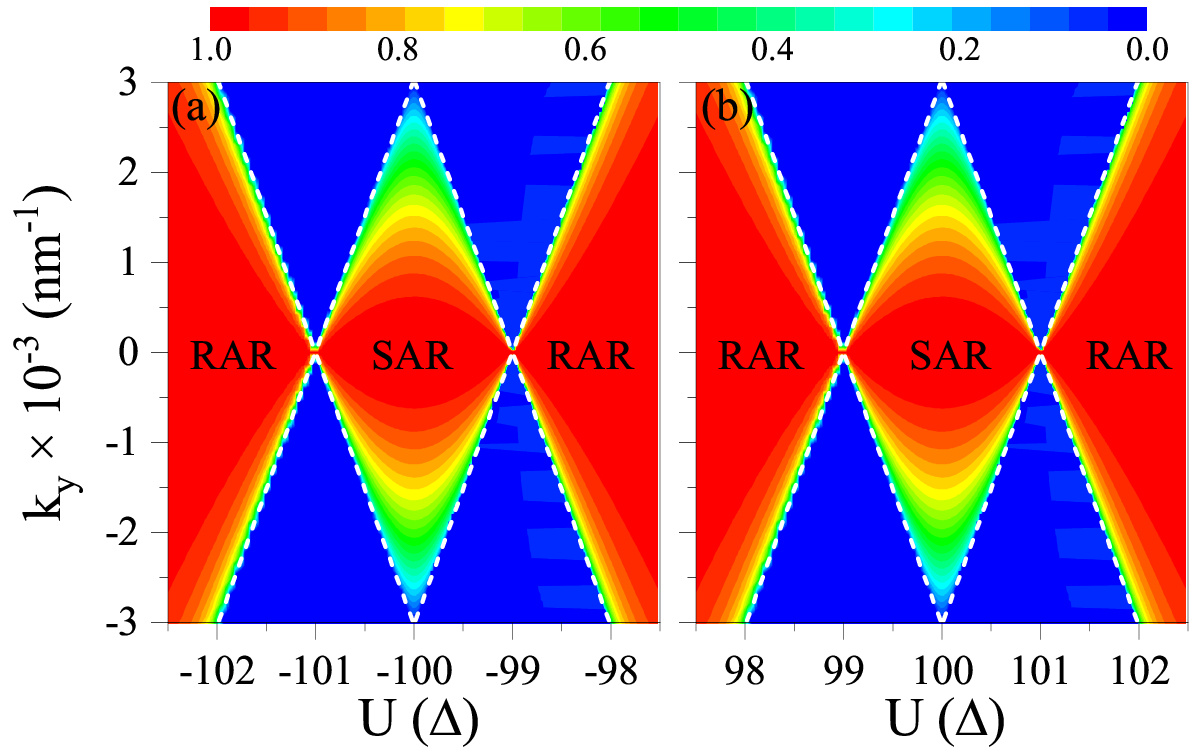}
\caption{ AR probabilities $R^{++}_A$ and $R^{--}_A$ in the $(U, k_y)$ plane for the incident electron from the (a) upper cone and (b) lower cone at $E=\Delta$ and $\delta=0$.
The white dashed curves are energy bands in the AABG region, which are also the boundaries of the zero and non-zero probabilities.}
\end{figure}

Fig. 2 shows the AR probabilities $R^{++}_A$ and $R^{--}_A$ in the $(U, k_y)$ plane for the incident electron from (a) upper cone and (b) lower cone at $E=\Delta$.
One can clearly see that the SAR could happen in the region $U_{c1}<U<U_{c2}$ for the upper cone.
When $\delta=0$, $U_{c1(c2)} = -\gamma -(+)E =-100\Delta -(+)E$. Namely, the SAR could happen in the region $|\gamma+U|<E$.
On the other hand, the AR is RAR in the region $|\gamma+U|>E$.
Similarly, for the lower cone, the SAR happens in the region $U_{c3}<U<U_{c4}$ (i.e., $|\gamma-U|<E$) and the RAR appears in the region $|\gamma-U|>E$, as shown in Fig. 2(b).
Here the SAR occurs between the conduction band and valence band in the intra-cone, and the RAR is in the intraband (the conduction band or valence band) in the intra-cone.
In particular, when the potential $U$ is in the region $-\gamma-E<U<-\gamma+E$, the AR processes are respectively the SAR and RAR
for the incident electron from the upper cone and lower cone [see Fig. 1(b)].
When $\gamma-E<U<\gamma+E$, the AR processes are RAR and SAR for the electron from the upper and lower cones [see Fig. 1(d)].
Therefore, based on the AR processes, different cone carriers can be spatially separated according to their cone indices by adjusting the potential $U$.
Here the potential $U$ can be well controlled by the gate voltage in the experiment.
In addition, because of the duplicate band structures between the two cones, the profiles of their ARs are the same, as exhibited in Figs. 2(a) and 2(b).
The AR probabilities are symmetric with respect to $k_y=0$ due to the invariance of the Hamiltonian $\tau_x H_\eta(k_y) \tau_x^{-1}=H_\eta(-k_y)$.
The SAR for the upper and lower cones are also symmetric about $U=-\gamma$ and $\gamma$ (here $\gamma=100\Delta$), because of the intra-cone particle-hole symmetry.
Furthermore, the AR probability is equal to one at $k_y=0$, indicating a complete conversion of electron to hole for the normal incidence case.

\begin{figure}
\includegraphics[width=8.0cm,height=5.0cm]{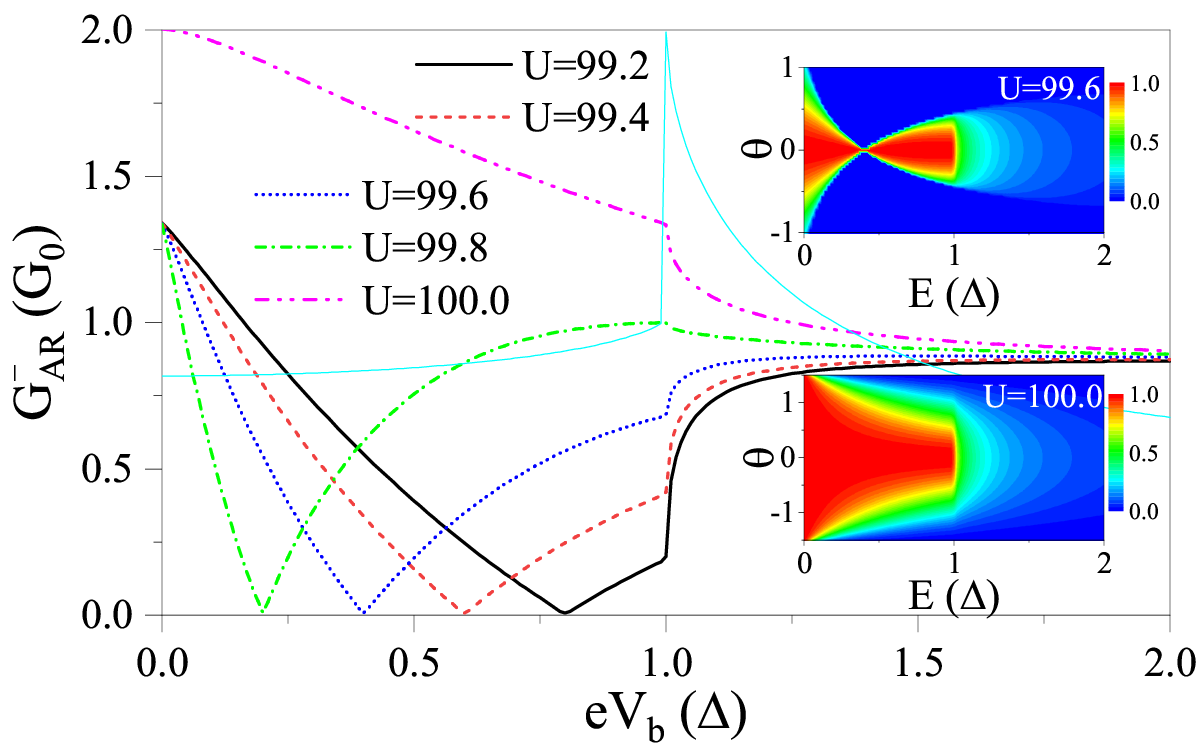}
\caption{ Andreev conductance $G^-_{AR}$ of the AABG/SC interface for the lower cone at $\delta=0$.
The insets show the AR probability $R^{--}_A$ in the $(E, \theta)$ plane at $U=99.6\Delta$ and $100.0\Delta$.
The cyan curve are the Andreev conductance $G^+_{AR}$ for the upper cone at $\delta=0$ and $U=99.4\Delta$.}
\end{figure}

Taking the lower cone for example, Fig. 3 shows the Andreev conductance $G^-_{AR}$ of the AABG/SC interface for different values of potential $U$ near the critical values $U_{c3(c4)}$.
The upper cone has the similar properties near the critical values $U_{c1(c2)}$.
In the subgap energy region, the AR is RAR when $eV_b<\gamma-U$ and it is SAR when $eV_b>\gamma-U$.
The conductance $G^-_{AR}$ contributed by RAR decreases with $eV_b$ and reduces to zero at $eV_b=\gamma-U$ where RAR crosses over to SAR.
At $eV_b=\gamma-U$, the mode number of the incident electron from the lower cone becomes zero, leading to the disappearance of the conductance.
Then the conductance $G^-_{AR}$ by SAR increases from zero with the further increase of $eV_b$.
When the potential increases to $U=100.0\Delta$, $G^-_{AR}$ is completely governed by SAR which decreases monotonously with $eV_b$.
The insets in Fig. 3 show the AR probability $R^{--}_A$ in the $(E, \theta)$ plane at $U=99.6\Delta$ and $100.0\Delta$.
As $E$ increases at $U=99.6\Delta$, the angle range for RAR probability reduces rapidly while the angle range for SAR probability becomes increasing.
At $U=100.0\Delta$, the SAR probability is weakened gradually with the increase of $E$.
The property of Andreev conductance can be understood by equation (18) and the AR probability in the $(E, \theta)$ plane.
In addition, in the limit $eV_b \rightarrow 0$ one has $G^-_{AR} \rightarrow 4G_0/3$ for RAR process. Such a characteristic is also proved in other systems \cite{Beenakker2}.
Fig. 3 also exhibits the Andreev conductance $G^+_{AR}$ for the upper cone at $\delta=0$ and $U=99.4\Delta$ (see the cyan curve).
We can see that $G_{AR}^{+}$ always has a large value while $G_{AR}^{-}$ is zero at $U=99.4 \Delta$ and $eV_b=\gamma-U=0.6\Delta$. 
Therefore, the Andreev conductance of the two cones can be separately measured near the critical values $U_{c1(c2)}$ and $U_{c3(c4)}$. 

\begin{figure}
\includegraphics[width=8.0cm,height=5.0cm]{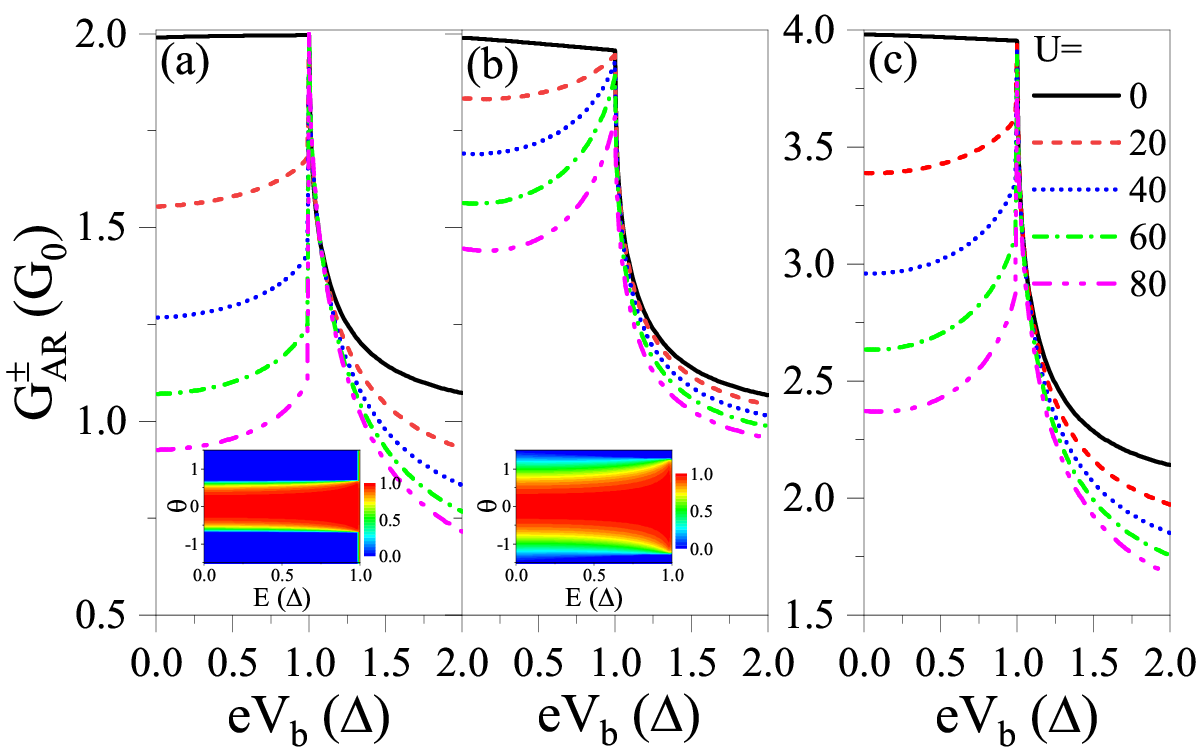}
\caption{ Andreev conductance (a) $G^{+}_{AR}$, (b) $G^{-}_{AR}$, and (c) $G_{AR}^{total}$ of the AABG/SC interface at $\delta=0$.
The insets in (a) and (b) show the AR probabilities $R^{++}_A$ and $R^{--}_A$ in the $(E, \theta)$ plane at $U=60.0\Delta$ and $\delta=0$, respectively.}
\end{figure}

Fig. 4 shows the property of Andreev conductance $G^{\pm}_{AR}$ in a wider range of potential $U$ for the (a) upper cone and (b) lower cone.
From Figs. 4(a) and 4(b), one may find that although the AR is RAR for both cones in the considered region, the features of the two RAR processes are different.
With the increase of $U$, the conductances $G^{\pm}_{AR}$ for both cones decrease gradually, but the decrease of $G^+_{AR}$ for upper cone is more significant.
$G^{-}_{AR}$ is larger than $G^{+}_{AR}$ for given $eV_b$ and $U$, because $R^{--}_A$ occurs in a wide angle range while $R^{++}_A$ occurs in a narrow angle range [see the insets in Figs. 4(a) and 4(b)], corresponding to the critical angles of Eq. (27) in the following. 
In the superconducting band gap, $G^{\pm}_{AR}$ increases with $eV_b$, and the increase is more significant for the lower cone, due to the increase behavior of $R^{++}_A$ and $R^{--}_A$ as functions of $E$.
The Andreev conductance $G^+_{AR}$ for upper cone has a maximum value at $eV_b=\Delta$ where $G^+_{AR}=2G_0$, regardless of the value of $U$.
However, the conductance $G^-_{AR}$ keeps declining as $U$ increases at $eV_b=\Delta$ for the lower cone.
Fig. 4(c) presents the total conductance $G_{AR}^{total}=G_{AR}^{+}+G_{AR}^{-}$, and its feature is similar to that of $G_{AR}^{+}$ and $G_{AR}^{-}$. 

\subsection{Double Andreev reflections}

\begin{figure}
\includegraphics[width=8.0cm,height=7.0cm]{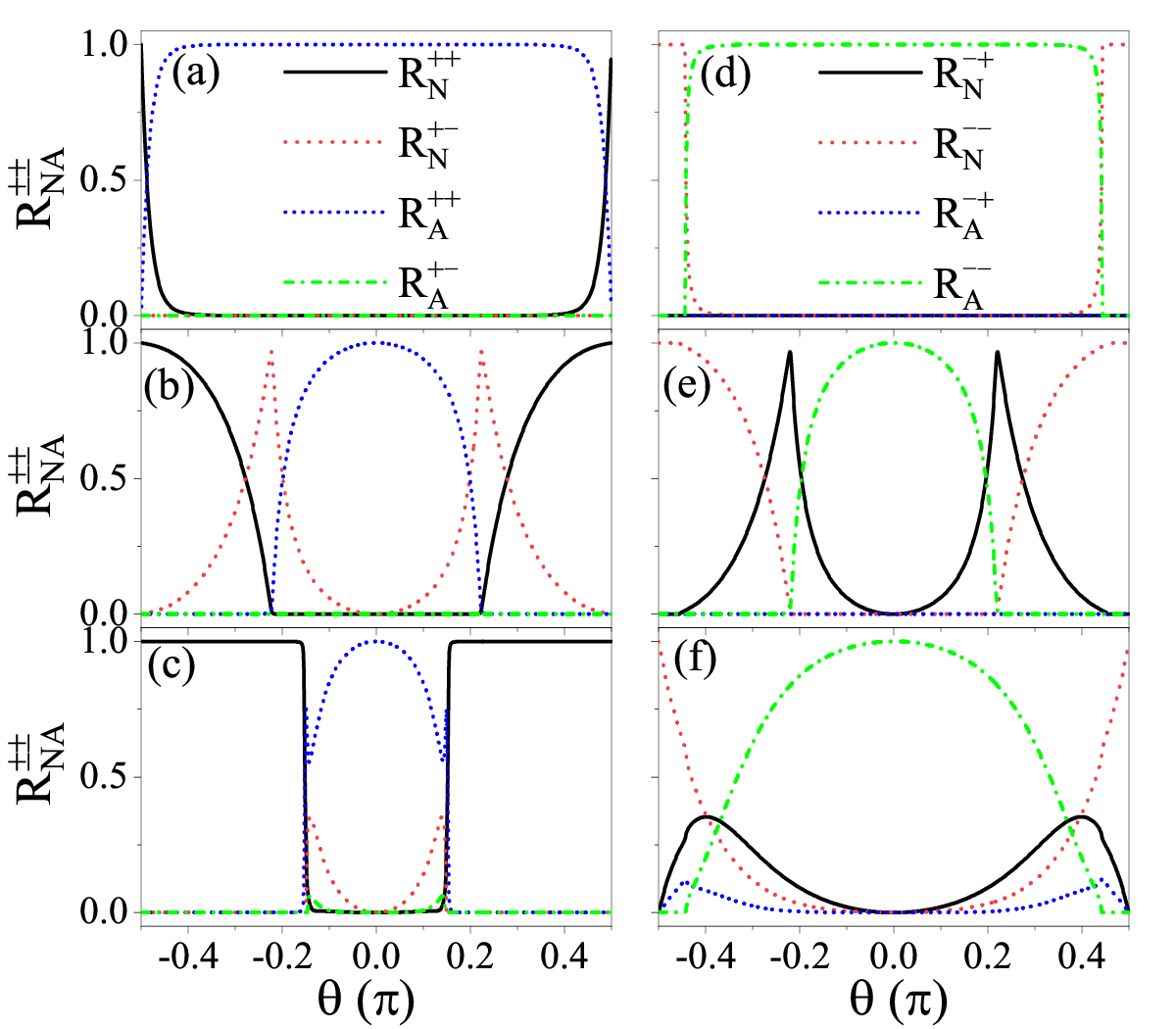}
\caption{ AR and NR probabilities $R^{\pm \pm}_{N,A}$ versus incident angle $\theta$ for the electron from the (a-c) upper cone and (d-f) lower cone at $E=0.8\Delta$.
The parameters $U=0.0$ and $\delta=0.0$ in (a) and (d), $U=0.0$ and $\delta=120.0\Delta$ in (b) and (e), and $U=60.0\Delta$ and $\delta=120.0\Delta$ in (c) and (f).}
\end{figure}

Next, we discuss the double ARs and double NRs at $\delta\neq0$ in which the inter-cone AR and inter-cone NR can occur, and the results are given in Figs. 5-8.
Fig. 5 presents the AR and NR probabilities $R^{\pm \pm}_{N,A}$ as functions of incident angle $\theta$ for the incident electron from the (a-c) upper cone and (d-f) lower cone.

When $U=\delta=0.0$ in Figs. 5(a) and 5(d), the inter-cone AR and inter-cone NR processes are forbidden, that is $R^{+-}_{N}=R^{-+}_{N}=R^{+-}_{A}=R^{-+}_{A}=0$.
Interestingly, the perfect intra-cone ARs with $R^{++}_{A}$ and $R^{--}_{A}$ almost being one appear in a wide angle region.
The relation between the incident angle $\theta$ and the AR angle $\phi$ satisfies $k_{e1(e2)}\sin\theta=k_{h1(h2)}\sin\phi$ due to the conservation of the transverse momentum.
When $\phi=\pi/2$, we can obtain the critical angles for the intra-cone ARs $R^{++}_A$ and $R^{--}_A$, $\theta^A_{c++(c--)}=\pm \arcsin [k_{h1(h2)}/k_{e1(e2)}]$, that is
\begin{align}
\theta^A_{c++(c--)}=\pm \arcsin[(E+U+(-)\Gamma)/(E-U-(+)\Gamma)]
\end{align}
for the upper cone ($c++$) and lower cone ($c--$).

When $U=0$ and $\delta \neq 0$ in Figs. 5(b) and 5(e), the inter-cone NR occurs, but the inter-cone AR is still forbidden.
In this situation, the total reflection has
\begin{align}
R^{++}_{N}+R^{+-}_{N}+R^{++}_{A} = R^{-+}_{N}+R^{--}_{N}+R^{--}_{A}= 1
\end{align}
for the upper and lower cones.
For both cones, the range of intra-cone ARs $R^{++}_A$ and $R^{--}_A$ are greatly reduced by $\delta$.
The critical angles for inter-cone NRs $R^{+-}_N$ and $R^{-+}_N$ can be obtained as
\begin{align}
    \theta^N_{c+-(c-+)}=\pm \arcsin[(E-U+(-)\Gamma)/(E-U-(+)\Gamma)].
\end{align}
$R^{+-}_N$ and $R^{-+}_N$ reach the maximum value near the incident angle $\theta=\pm 0.22\pi$ and show the peaks.
The intra-cone NRs $R^{++}_N$ and $R^{--}_N$ reach the maximum value at the angle $\theta=\pm\pi/2$ and they monotonously reduce with the decrease of the angle $|\theta|$.

As predicted by Eq. (21), when $U \neq 0$ and $\delta \neq 0$ in Figs. 5(c) and 5(f), not only inter-cone NR but also inter-cone AR could occur.
Thus, double ARs and double NRs take place simultaneously in the AABG/SC junction.
The critical angles for the inter-cone ARs $R^{+-}_A$ and $R^{-+}_A$ are
\begin{align}
    \theta^A_{c+-(c-+)}=\pm \arcsin[(E+U-(+)\Gamma)/(E-U-(+)\Gamma)].
\end{align}
For incident electron from the upper cone, the inter-cone AR $R^{+-}_A$ and NR $R^{+-}_N$ mainly occur near the critical angle $\theta^A_{c+-} \approx \pm 0.15 \pi$ [see Fig. 5(c)].
The intra-cone AR $R^{++}_A$ also mainly appears in the range of  ($-\theta^A_{c+-}$, $\theta^A_{c+-}$), and outside of this range $R^{++}_A$ is very small but not zero. The intra-cone NR $R^{++}_N$ mainly occurs in the range from $\pm\pi/2$ to $\pm\theta^A_{c+-}$ [see Fig. 5(c)].
For incident electron from the lower cone, the inter-cone AR $R^{-+}_A$ and inter-cone NR $R^{-+}_N$ can occur in the range from $-\pi/2$ to $\pi/2$, and they are larger in the vicinity of $\pm\pi/2$ [see Fig. 5(f)].
In addition, the intra-cone AR $R^{--}_A$ can occur and keep the large value at a wide angle range [see the green dotted-dashed curve in Fig. 5(f)].
Note that the intra-cone ARs $R^{--}_A$ and $R^{++}_A$ always keep perfect reflection at the normal incidence ($\theta=0$).

\begin{figure}
\includegraphics[width=8.0cm,height=3.0cm]{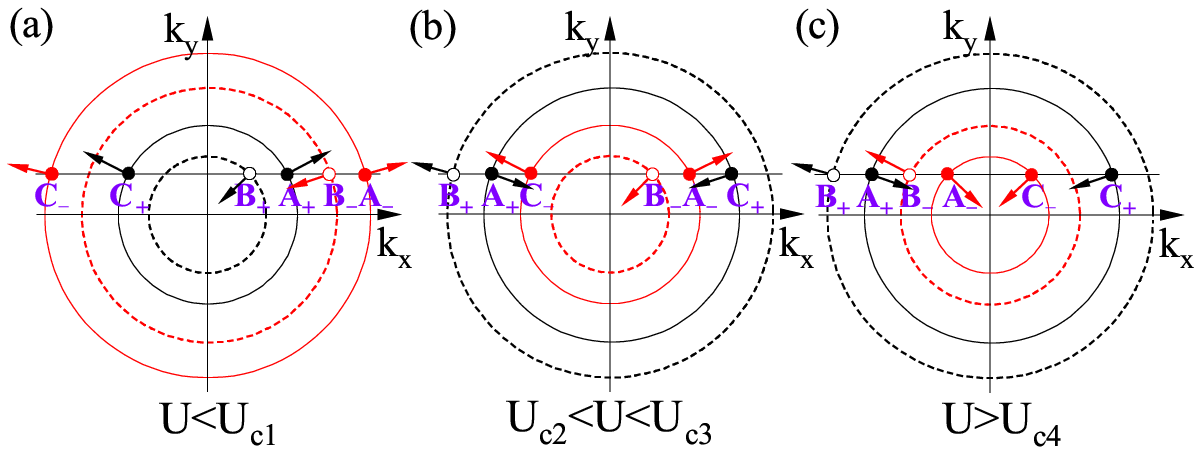}
\caption{ The projection of the Andreev scattering and normal scattering processes on the $k_x-k_y$ plane when (a) $U<U_{c1}$, (b) $U_{c2}<U<U_{c3}$, and (c) $U>U_{c4}$. The black and red arrows denote the group velocity at upper and lower cones.
The arrows with filled circles and empty circles are for the electrons and holes, respectively.}
\end{figure}

Based on Eqs. (4), (5), (22) and (23), we can classify the ARs and NRs by adjusting $U$ and $\delta$, as shown in Figs. 1(b), 1(d), and 6.
One may find that for the intra-cone process, the SAR for the incident electron from upper cone (or lower cone) only appears when $U_{c1}<U<U_{c2}$ (or $U_{c3}<U<U_{c4}$) [see Figs. 1(b) and 1(d)].
The intra-cone NR is always SNR, independent of the value of $U$.
For the inter-cone process, the SAR for the electron from upper cone could occur when $U_{c2}<U<U_{c3}$ [see Fig. 6(b)], otherwise, the inter-cone AR is RAR.
The inter-cone SAR for the electron from lower cone could happen in a broader region $U_{c1}<U<U_{c4}$ [see Figs. 1(b), 1(d), and 6(b)].
Note that the potential range $U_{c2}<U<U_{c3}$ or $U_{c1}<U<U_{c4}$ for the appearance of the inter-cone SARs is about $2\Gamma =2\sqrt{\gamma^2+\delta^2}\geq 2\gamma$.
This potential range is a few hundred of $meV$, which is much larger than the potential range for the intra-cone SARs, a few of superconducting gap $\Delta$.
Dramatically, the inter-cone RNR, an abnormal NR, would occur when $U_{c2}<U<U_{c4}$ for both cones [see Figs. 1(d) and 6(b)], where the path of reflected electron is the same as that of incident electron [see Fig. 1(c)].
Therefore, we can realize a spatially separated cone carrier by regulating the potential $U$ in the AABG/SC junction.

\begin{figure}
\includegraphics[width=8.0cm,height=6.0cm]{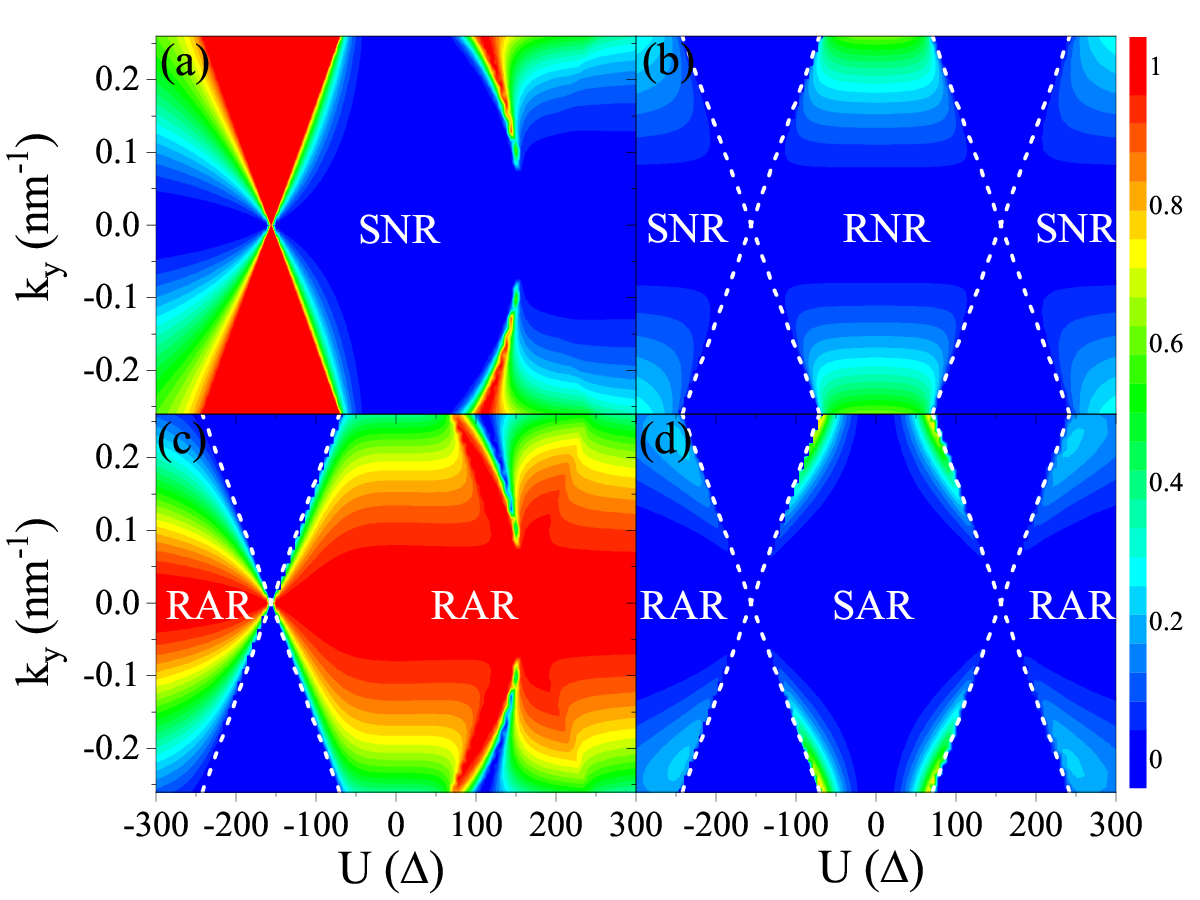}
\caption{ AR and NR probabilities (a) $R^{++}_{N}$, (b) $R^{+-}_{N}$, (c) $R^{++}_{A}$, and (d) $R^{+-}_{A}$ in the $(U, k_y)$ plane for the electron from the upper cone at $E=0.0$ and $\delta=120.0\Delta$.
The white dashed curves are energy bands in the AABG region, which are also the boundaries of the zero and non-zero probabilities.}
\end{figure}

Fig. 7 displays double ARs and double NRs in the $(U, k_y)$ plane for the electron from upper cone when $E=0.0$.
The distributions of various types of AR and NR in Fig. 7 are consistent with the results in Figs. 1(b), 1(d), and 6.
The intra-cone reflections take a leading role in the scattering process [see Figs. 7(a) and 7(c)].
As expected, all the AR and NR are symmetric about $k_y=0$.
The inter-cone AR and NR are also symmetric about $U=0$, and they mainly appear in the range $|k_y|>0.1nm^{-1}$ [see Figs. 7(b) and 7(d)].
Similar results may be obtained for the lower cone.

\begin{figure}
\includegraphics[width=8.0cm,height=5.0cm]{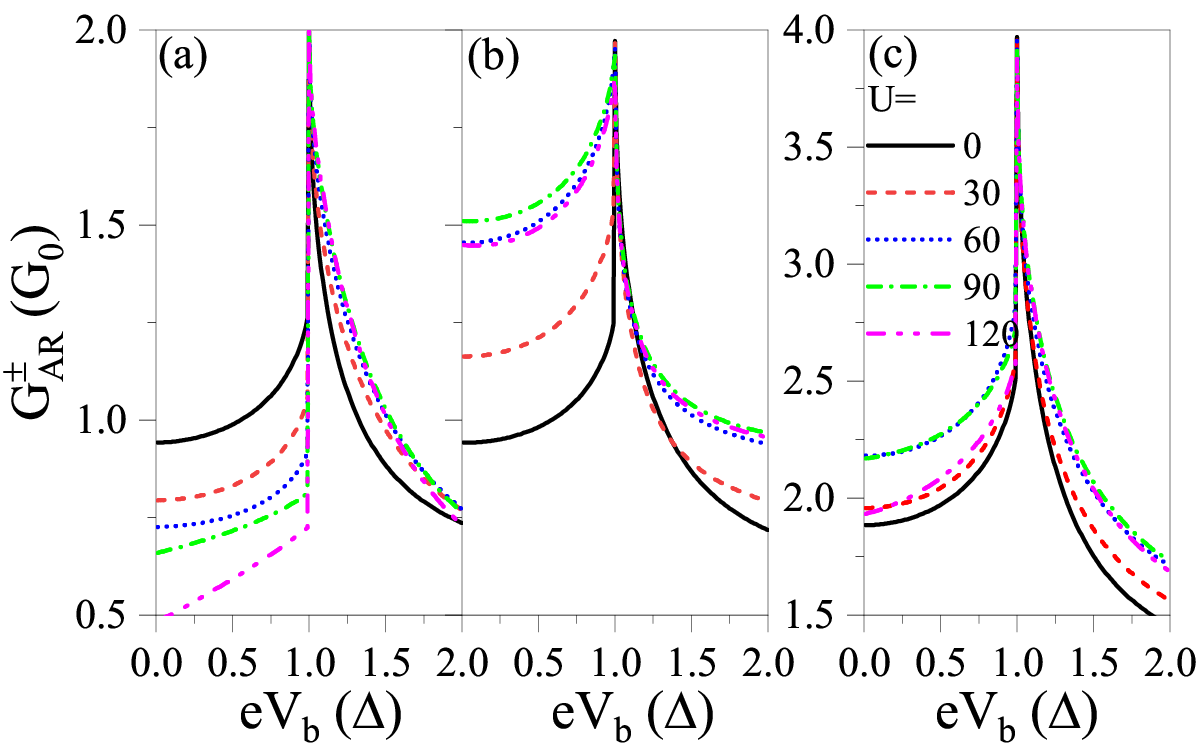}
\caption{ Andreev conductance (a) $G^{+}_{AR}$, (b) $G^{-}_{AR}$, and (c) $G_{AR}^{total}$ of the AABG/SC interface at $\delta=120.0\Delta$.}
\end{figure}

Finally, we discuss the Andreev conductance $G^{\pm}_{AR}$ for different values of the potential $U$ when double ARs and double NRs take place, as shown in Fig. 8, which is different from the one exhibited in Fig. 4.
Below we focus on the energy in the superconducting gap.
For the electron from upper cone, the conductance $G^+_{AR}$ decreases as $U$ increases, but $G^+_{AR}$ increases with the continuous increase of $eV_b$.
$G^+_{AR}$ presents an approximately linear change with $eV_b$ for a large $U$ such as $U=120.0\Delta$ [see Fig. 8(a)].
For the lower cone, the conductance $G^-_{AR}$ is a nonmonotonic function of the potential $U$. With the increase of $U$, $G^-_{AR}$ increases first and gets its maximum at $U \approx 90.0\Delta$, then decreases gradually [see Fig. 8(b)].
$G^{-}_{AR}$ remains larger than $G^{+}_{AR}$ for given $eV_b$ and $U$, since $R^{-\pm}_A$ appear in a wide angle range but $R^{+\pm}_A$ appear in a narrow angle range [see Figs. 5(c) and 5(f)], corresponding to the critical angles of Eqs. (27) and (30).
The conductance $G^{\pm}_{AR}$ for both the upper and lower cones reaches the maximum at $eV_b=\Delta$ and then it gradually decreases as $eV_b$ further increases.
Fig. 8(c) indicates that because of the occurrence of inter-cone AR and inter-cone NR when $\delta\ne 0$, the total conductance $G_{AR}^{total}$ is increased by the finite potential $U$, different from that at $\delta=0$. 

\section{Conclusion}

In summary, we studied the cone-dependent SAR and RAR in the AABG/SC junction by solving the BdG equation.
Due to the peculiar band structure of AABG, some fascinating scattering phenomena are found, such as double ARs, double NRs, inter-cone AR, inter-cone NR, which are not observed in monolayer graphene and other layered materials.
The inter-cone AR can be SAR and RAR, and the inter-cone SAR can exist in a large range of the potential energy.
The properties of AR and NR processes can be effectively controlled by regulating the potential and the inter-layer potential difference.
These results provide an electrical method for testing SAR in experiment and suggest a potential application in cone-tronic devices.

\section*{Acknowledgments}

We would like to thank Q. Cheng for many helpful discussions.
This work was supported by the National Natural Science Foundation of China (Grants No. 11974153, No. 11921005, and No. 12374034), the Innovation Program for Quantum Science and Technology (Grant No. 2021ZD0302403), and the Strategic Priority Research Program of Chinese Academy of Sciences (Grant No. XDB28000000).

\section*{Appendix: Eigenvectors of BdG Hamiltonian}
In this Appendix, we calculate the eigenvectors of BdG Hamiltonian.
For given energy $E$ and transverse momentum $k_y$, the eigenvectors in the AABG region have the form:
\begin{align}
\psi_{e1(2)}^{\pm} (x) =\frac{e^{\pm i k_{e1(2)x}x}}{\Lambda_{e1(2)}}  \left(\begin{array}{cc}    \varphi_{e1(2)}^{\pm}     \\    O       \end{array}\right),
\end{align}
\begin{align}
\psi_{h1(2)}^{\pm} (x) =\frac{e^{\pm i k_{h1(2)x}x}}{\Lambda_{h1(2)}}  \left(\begin{array}{cc}  O   \\   \varphi_{h1(2)}^{\pm}     \end{array}\right),
\end{align}
with
\begin{align}
\varphi_{e1(2)}^{\pm} = \left(\begin{array}{cc}    \hbar v_F[\pm k_{e1(2)x} - i k_y] [\delta+(-)\Gamma]    \\      +[E-U-(+)\Gamma] [\delta+(-)\Gamma]    \\  \hbar v_F[\pm k_{e1(2)x} - i k_y]\gamma     \\     +[E - U - (+) \Gamma ] \gamma    \end{array}\right),
\end{align}
\begin{align}
\varphi_{h1(2)}^{\pm} (x) = \left(\begin{array}{cc}    \hbar v_F[\pm k_{h1(2)x} - i k_y] [\delta+(-)\Gamma]  \\  -[E+U+(-)\Gamma] [\delta+(-)\Gamma]  \\  \hbar v_F[\pm k_{h1(2)x} - i k_y]\gamma  \\   -[E+U+(-)\Gamma]\gamma     \end{array}\right),
\end{align}
and $O=(0,0,0,0)^T$ is the null matrix.
Here, $\psi_{e(h)1}$ and $\psi_{e(h)2}$ are the wave functions for the electron (hole) near the upper and lower cones, respectively. The related parameters are defined as
\begin{align}
\Lambda_{e1(2)} = [E-U-(+)\Gamma] \sqrt{2[(\delta+(-)\Gamma)^2 + \gamma^2]},   
\end{align}
\begin{align}
\Lambda_{h1(2)} = [E+U+(-)\Gamma] \sqrt{2[(\delta+(-)\Gamma)^2 + \gamma^2]},
\end{align}
\begin{align}
k_{e1(2)x} = sgn [k_{e1(2)}] \sqrt{k_{e1(2)}^2 - k_y^2},   
\end{align}
\begin{align}
k_{h1(2)x} = sgn [k_{h1(2)}] \sqrt{k_{h1(2)}^2 - k_y^2},
\end{align}
\begin{align}
k_{e1(2)}=[E-U-(+)\Gamma]/\hbar v_F, 
\end{align}
\begin{align}
k_{h1(2)}=[E+U+(-)\Gamma]/\hbar v_F, 
\end{align}
$\Lambda_{e1(2)}$ and $\Lambda_{h1(2)}$ are the normalization factors for the electron and hole.
$k_{e1(2)x}$ and $k_{h1(2)x}$ are $x$ components of momentums $k_{e1(2)}$ and $k_{h1(2)}$, respectively.
The two eigenvectors in the SC region are given as
\begin{align}
\psi_{S1(2)}^{\pm} (x) =\frac{e^{\pm i q_{1(2)} x}}{\Lambda_{S1(2)}}  \left(\begin{array}{cc} \Omega [\Omega-(+)\gamma]     \\     \hbar v_F[\pm q_{1(2)} + i k_y]E  \\   +(-)E [\Omega-(+)\gamma]    \\     +(-)\hbar v_F[\pm q_{1(2)} + i k_y] \Omega  \\    0    \\   \hbar v_F[\pm q_{1(2)} + i k_y]\Delta   \\    +(-)\Delta [\Omega-(+)\gamma] \\  0  \end{array}\right),
\end{align}
\begin{align}
\psi_{S3(4)}^{\pm} (x) =\frac{e^{\pm i q_{1(2)} x}}{\Lambda_{S1(2)}}  \left(\begin{array}{cc} \hbar v_F[\pm q_{1(2)} - i k_y]E  \\  \Omega [\Omega-(+)\gamma]    \\    +(-)\hbar v_F[\pm q_{1(2)} - i k_y] \Omega  \\    +(-)E [\Omega-(+)\gamma]    \\     \hbar v_F[\pm q_{1(2)} - i k_y]\Delta   \\   0    \\  0   \\     +(-)\Delta [\Omega-(+)\gamma] \end{array}\right),
\end{align}
with
\begin{align}
\Lambda_{S1(2)}=2 E [\Omega-(+)\gamma],    
\end{align}
\begin{align}
q_{1(2)} = sgn [E-(+)\sqrt{\Delta^2+\gamma^2}] \sqrt{\frac{[\Omega-(+)\gamma]^2}{(\hbar v_F)^2} - k_y^2}, 
\end{align}
\begin{align}
\Omega = \sqrt{E^2-\Delta^2}.
\end{align}
$\Lambda_{S1(2)}$ is the normalization factor and $q_{1(2)}$ is $x$ component of momentum.

\end{CJK*}
\end{document}